\documentclass[english,keywords,amsmath,amssymb, twocolumn,showpacs]{revtex4}
\usepackage[T1]{fontenc}
\usepackage[latin1]{inputenc}
\usepackage{babel}%
\usepackage{graphicx}
\usepackage{color}
\usepackage{bm}
\usepackage{longtable}
\usepackage{amsmath}
\usepackage{amsfonts}
\usepackage{amssymb}

\begin{document}
\title{Cat state, sub-Planck structure and weak measurement}
\author{A. K. Pan$^{1,2}$ \footnote{akp@math.cm.is.nagoya-u.ac.jp}}
\author{P. K. Panigrahi$^{2}$\footnote{pprasanta@iiserkol.ac.in}}
\affiliation{$^{1}$Graduate School of Information Science, Nagoya University, Chikusa-ku, Nagoya 464-8601, Japan}
\affiliation{$^{2}$Indian Institute of Science Education and Research Kolkata, Mohanpur, Nadia 741252, India}
\begin{abstract}
Heisenberg-limited and weak measurements are the two intriguing notions, used in recent times for enhancing the  sensitivity of measurements in quantum metrology. Using a quantum cat state, endowed with sub-Planck structure, we connect these two novel concepts. It is demonstrated that these two phenomena manifest in complementary regimes, depending upon the degree of overlap between the mesoscopic states constituting the cat state under consideration. In particular, we find that when sub-Planck structure manifests, the imaginary weak value is obscured and vice-versa.   
\end{abstract}
\maketitle
\section{introduction}
In quantum mechanics (QM), measurement plays a fundamental role, because the precession of the same is, not only dependent on technology, but also on the inherent fundamental constraints imposed by the theory itself. 
The Heisenberg-limited \cite{zurek} and the weak measurements \cite{aav} are the two notions that are being currently used for enhancing the sensitivity of measurements. 

The Sub-Planck phase-space structures are at the root for the Heisenberg-limited sensitivity, accessible through some specific quantum states, e.g., the cat state and its generalizations. It explains oscillating photon number distribution of squeezed states in a phase plane \cite{schl}. The Wigner distribution ideally captures such interference phenomena, as it can be negative in a phase-space region \cite{intf, vogel}. In a thought-provoking work, using a compass state (superposition of four suitable Gaussian states), Zurek first demonstrated \cite{zurek} its relevance for Heisenberg-limited sensitivity \cite{toscano} for certain parameter estimation. Later, a classical wave optics analogue has experimentally been tested\cite{pryx}. For quasiclassical states the sensitivity is restricted by the standard quantum limit, also known as the shot-noise limit. Sub-shot-noise sensitivities can be obtained using the sensitivity of the quantum state to displacements, which is related to the sub-Planck phase space structure \cite{roy}. A number of proposals have been advanced for generating single particle cat and generalized states, showing the above feature \cite{toscano,gsa}.

The path-breaking idea of weak measurement (WM) in QM, originally proposed by Aharonov, Albert and Vaidman (AAV) \cite{aav}, has gained wide interest in realizing apparently counterintuitive quantum effects. In this measurement scenario, the empirically measured value (coined as
`weak value') of an observable can be seemingly puzzling, in that, it yields results going beyond the eigenvalue spectrum of the measured observable. This idea has been further enriched by a number of theoretical \cite{duck,av91,mit,jeff,jozsa,brunner,panmatz} and experimental works \cite{ritchie,wang,pryde,hosten,yokota,starling,lundeen09,lund,zil}. WM has several implications, the primary one being that, it provides insight into conceptual quantum paradoxes \cite{av91,vaidman,aha02,mit,geo,matzin}. At a practical level, it helps in identifying tiny spin Hall effect \cite{hosten}, detecting very small transverse beam deflections \cite{starling}, measuring average quantum trajectories for photons \cite{steinberg11} and protecting a quantum state\cite{kim}.

Both the sub-Planck structure and the weak measurement are striking quantum mechanical phenomena, which rely on the interference effect.  In this paper, by establishing an interesting connection between these two seemingly different illustrations of interference effect, we show that they are in fact complementary to each other. Using a cat state, we demonstrate that the parameter regime relevant for Heisenberg limited measurement is opposite to what is required for weak measurement. 

For our purpose, we consider a cat state, a superposition of two Gaussian wave functions in position representation, 
\begin{align}
\label{cat}
	\Psi_{c}(x)= \mathcal{N}\left(a \psi_{+}(x)+ b \psi_{-}(x)\right), 
\end{align}
where $a$ and $b$ are in general complex, satisfying $|a|^{2}+|b|^{2}=1$. The normalization constant $\mathcal{N}=1/\sqrt{(1+ 2 I \Re[a^{\ast}b])}$  where $I=\int \psi_{+}^{\ast}(x)\psi_{-}(x) dx$.

The wave function $\psi_{\pm}(x)$ are taken to
be  Gaussian, peaked at ${x}=\pm x_{0}$ respectively, given by,
\begin{align}
\label{catpm}
	\psi_{\pm}(x)=\frac{1}{\left(  2\pi\eta^{2}\right)
^{\frac{1}{4}}} exp\left[-\frac{(x \mp x_{0})^{2}}{4 \eta^{2}} \pm \ i \frac{p_{0} x}{\hbar}\right].
\end{align}
where $\eta$ is the initial width of the wave packet. The wave packets corresponding $\psi_{\pm}(x)$  move along the $\pm x$ axis with the initial momentum $p_{0}$. The inner product $(I)$ between 	$\psi_{\pm}(x)$ is given by 
\begin{equation}
\label{ip}
I= e^{- \frac{2 p_{0}^{2} \eta^{2}}{ \hbar^{2}} - \frac{x_{0}^{2}}{2 \eta^{2}}}.
\end{equation} 
which plays a crucial role in this paper. As is obvious, the value of $I$ ranges from $0$ to $1$. In the following, it will be shown that the sub-Planck structure emerges at one extreme limit, when $I\approx0$, and the weak value is obtained at another extreme limit, when $I\approx1$, i.e., the same quantum system can reveal both the fundamental effects in two opposite regimes. More explicitly, we demonstrate that, when there is an existence of weak value, the sub-Planck structure is obscured and vice-versa. We will see later in the Appendix that $I\approx0$ and $I\approx1$, respectively imply the strong and weak measurement scenarios, when we show $\Psi_{c}(x)$ as meter state after post-selection in AAV weak measurement setup.

The paper is organized as follows. In Sec. II, we recapitulate the essences of sub-Planck structure and weak measurement and highlight the crucial role of interference in both of the cases. Sec. III is devoted to illustrations that connect these two concepts. We conclude in Sec. IV, and point out a number of directions for future works. An Appendix is also included explaining the procedure for generating the cat state by using a series of suitable Stern-Gerlach setups, employed for AAV weak measurement of a dichotomic observable, which is actually the post-measurement final state allowing one to infer the weak value.

\section{sub-Planck structure and weak value associated with cat state}
As mentioned earlier, sub-Planck structure plays an important role in distinguishing quantum states, slightly displaced from each other, as was first demonstrated for a compass state \cite{zurek}. It was conjectured there that, it can also manifest in cat state, which was later proved \cite{pryx,toscano} - the effect being known as  sub-Fourier sensitivity, a special case of sub-Planck one. Instead of using Wigner function, the sub-Fourier feature can be demonstrated in the following way. If a superposition of two Gaussian wave functions is subjected to a sub-Fourier shift in phase space, then surprisingly the shifted one can be orthogonal to the original one, i.e., the shifted and the original ones are distinguishable and the small sub-Fourier shift can be measured. Hence, the determination of the sub-Planck phase shift is the key in this scheme. 

Explicitly, let the cat state is phase shifted by an amount $\delta$, so that the resulting state is given by, $|\Psi^{\prime}_{c}\rangle=	|\Psi_{c}\rangle e^{i x\delta}$. The overlap between the shifted and the original states can be obtained as,
\begin{eqnarray}
\label{ov}
\nonumber
&&|\langle \Psi^{\prime}_{c}	|\Psi_{c}\rangle|^{2}= \mathcal{N}^4 e^{-\delta^2 \eta^2}[|a|^{4}+ |b|^{4} +  2 |a|^{2}|b|^{2} cos(2 x_{0}\delta) \\
\nonumber
&+&2 I \cos(x_{0}\delta) \left(a^{\ast}b e^{z} + b^{\ast}a e^{-z}\right)\\
&+& I^2 \left\{(a^{\ast})^{2} b^{2} e^{2z} + a^2 (b^*)^{2} e^{-2z} + 2|b|^2 |a|^{2}\right\}],
\end{eqnarray}
 with $z=2  p_{0}\eta^{2}\delta/\hbar$.

Note here that, given the cat state and other relevant parameters $x_{0}$, $p_0$ and $\eta$, the overlap function vanishes for multiple values of  $\delta$, enabling one to carry out Heisenberg limited measurements. It is seen from Eq.(\ref{ov}), that when $I\approx0$, the last two terms give no contribution - the case when the distance between two nearest zeros is the minimum, allowing one to carry out the most sensitive measurement of the shift parameter $\delta$. For $I\approx1$, the distance between two nearest zeros tends to its maximum value and sub-Planck structure gets obscured. 

Now, the associated Wigner function can be written as

\begin{eqnarray}
&&W(x,p)=\mathcal{N}^2 exp\left(-\frac{(x + x_{0})^2}{2 \eta^2} - \frac{  2 (p + p_{0})^2 \eta^2}{\hbar^2}\right) \\
\nonumber
&\times&
[ a a^{\ast} exp\left(\frac{2 x x_{0}}{\eta^2} + \frac{8 p p_{0}\eta^2}{\hbar^2}\right) + b b^{\ast}\\
\nonumber
&+& exp\left(\frac{x_{0} (2 x + x_{0})}{2 \eta^2} + \frac{     2 p_0 (2 p + p_0) \eta^2}{\hbar^2}\right) \\
\nonumber
&& \times \left\{Re(a^{\ast} b) cos\frac{2 p_0 x - 2 p \ x_{0}}{\hbar}-Im(a^{\ast} b)\sin\frac{2 p_0 x -2 p x_{0}}{\hbar }\right\}]
\end{eqnarray}

which reveals the signature of the interference in phase space, if any. 

Next, we demonstrate how the weak value emerges from the same cat state. In contrast to strong measurement, the formalism of weak measurement allows one to extract information of a quantum system in the limit of vanishingly small disturbance. While measuring an observable in strong measurement, the pointer indicates the eigenvalues of the given observable. In weak measurement, the pointer may indicate a value (the weak value) beyond the eigenvalues range. Let a system be prepared in a state (in popular terminology, pre-selected state)  $|\chi_{i}\rangle$ and $\hat{O}$ being the observable to be measured. If the measurement interaction between the system and the meter is weak, the system state remains grossly unchanged. The most interesting step in weak measurement is the post-selection in a state, say, $|\chi_{f}\rangle$ after the intermediate interaction. After post-selection, the meter will be left in a final state allowing to access the weak value.  The standard definition of the weak value is given by,
\begin{equation}
(O)_{w}=\frac{\left\langle \chi_{f}|O|\chi
_{in}\right\rangle }{\left\langle \chi_{f}\right\vert \chi_{in}\rangle
}\label{aavwvalue},%
\end{equation}
which can be widely outside the eigenvalues range and, can even be complex. For example, if $|\chi_{in}\rangle= a_1 |\uparrow_{z}\rangle+ a_2|\downarrow_{z}\rangle$, $|\chi_{f}\rangle=\left\vert \uparrow
\right\rangle _{z}$ and the observable $\hat{O}= \hat{\sigma}_{x}$, then the weak value is given by
$(\sigma_{x})_{w}=a_{2}/a_{1}$. Hence, the value of $(\sigma_{x})_{w}$ can become arbitrarily large as $a_{1}\rightarrow 0$ \cite{aav}. Since $a_1$ and $a_{2}$ are in general complex satisfying $|a_{1}|^{2} + |a_{2}|^{2}=1$, the weak value can be complex and large. If the interaction is defined in position space and initial pointer wave function is Gaussian, the real and imaginary parts of the weak value produce the pointer shifts in momentum and position spaces respectively. In this paper, we shall use the imaginary eccentric weak value and look at the shift in position distribution.  

We now consider the cat state given by Eq.(\ref{cat}) and find the weak value associated with it. Let this state be the post-selected final pointer state, obtained by post-selecting the system in the state $|\chi_{f}\rangle$ by identifying $a= a_{1}+ a_{2}$ and $b=a_{1}- a_{2}$ if the relevant pre-selected and dichotomic observable in question are taken to be $|\chi_{in}\rangle= a_1 |\uparrow_{z}\rangle+ a_2|\downarrow_{z}\rangle$ and $\hat{O}= \hat{\sigma}_{x}$ respectively. The detail procedure of obtaining such a cat state in relation to weak measurement is given in the Appendix. This state is expected to point at the weak value, which can be demonstrated as follows.  

For this, we consider the situation when the value of $I$ is close to unity which can be obtained for  very small values of $x_{0}$ and $p_{0}$ for a given value of $\eta$. (In the Appendix, we provided a detailed description of weak measurement setup where we discussed that $I\approx 1$ implies the weak measurement of the observables $\hat \sigma_{x}$.) In such a case, we can neglect $x_{0}$ dependency in the amplitude part of $\psi_{\pm }\left(  {x}\right)$ and the exponential of $\psi_{\pm }\left(  {x}\right)$ can be expanded up to the first order of $p_{0}$:
\begin{equation}
\psi_{\pm }\left(  {x}\right)  \approx e^{-\frac{x ^2}{4 \eta^{2}}} \left(  1\pm
ip_{0}x/\hbar\right) \label{appwfn}.
\end{equation}  
Note that, we skipped the normalization constant here which merely adds the weight. In this limit Eq.(\ref{cat}) becomes
\begin{equation}
\label{ppp}
\Psi_{c}({x})\approx {\left(
2\pi\eta^{2}\right)  ^{1/4}} e^{-\frac{x ^2}{4 \eta^{2}}}\left[  a\left(1+ip_{0} x/\hbar\right)  +b\left(  1-ip_{0}x/\hbar\right)  \right].
\end{equation}
Substituting the values of $a$ and $b$, and simplifying one gets
\begin{equation}
\label{psii1}
\Psi_{c}(x)\approx 2 a_1 {\left(
2\pi\eta^{2}\right)  ^{-1/4}}\exp\left(  -\frac{x^{2}}{4\eta^{2}}%
+i\frac{p_{0} x (a_{2}/a_{1})}{\hbar}\right).
\end{equation}
Now, if the weak value( $(\hat\sigma_{x})_{w}=a_2/a_1$) is imaginary, $|\Psi_{c}(x)|^{2}$ gives the pointer to be peaked at the weak value. In this work, we shall be using the imaginary weak value. However, if the weak value is real, one has to look at the momentum probability distribution is of the form 
\begin{equation}
|\Phi_{c}(p_{x})|^{2}\approx 4 |a_1|^{2} \left(  \frac{2\eta^{2}%
}{\pi\hbar^{2}}\right)  ^{1/2}\exp\left[  -\frac{2\eta^{2} (p_{x}
-p_{0}(a_{2}/a_{1}))^{2}}{\hbar^{2}}\right],
\end{equation}
Note again here that, this argument is valid for weak coupling regime, implying the inner product ($I$) between $\psi_{\pm }\left(  {x}\right)$ is close to unity, as given in the Appendix.  

Thus, the cat state, in one hand gives the sub-Planck structure and on the other hand provides pointer shift corresponding to the weak value of a dichotomic observable. We will now examine, whether the sub-Planck structure and weak value pointer shift can be obtained simultaneously for the same parameter choices. We will see in the next section that, when $I\approx0$, there is no weak value and when $I\approx1$, there is no interference in phase space yielding no sub-Planck structure.
\begin{figure}[h]
{\rotatebox{0}{\resizebox{6.0cm}{5.0cm}{\includegraphics{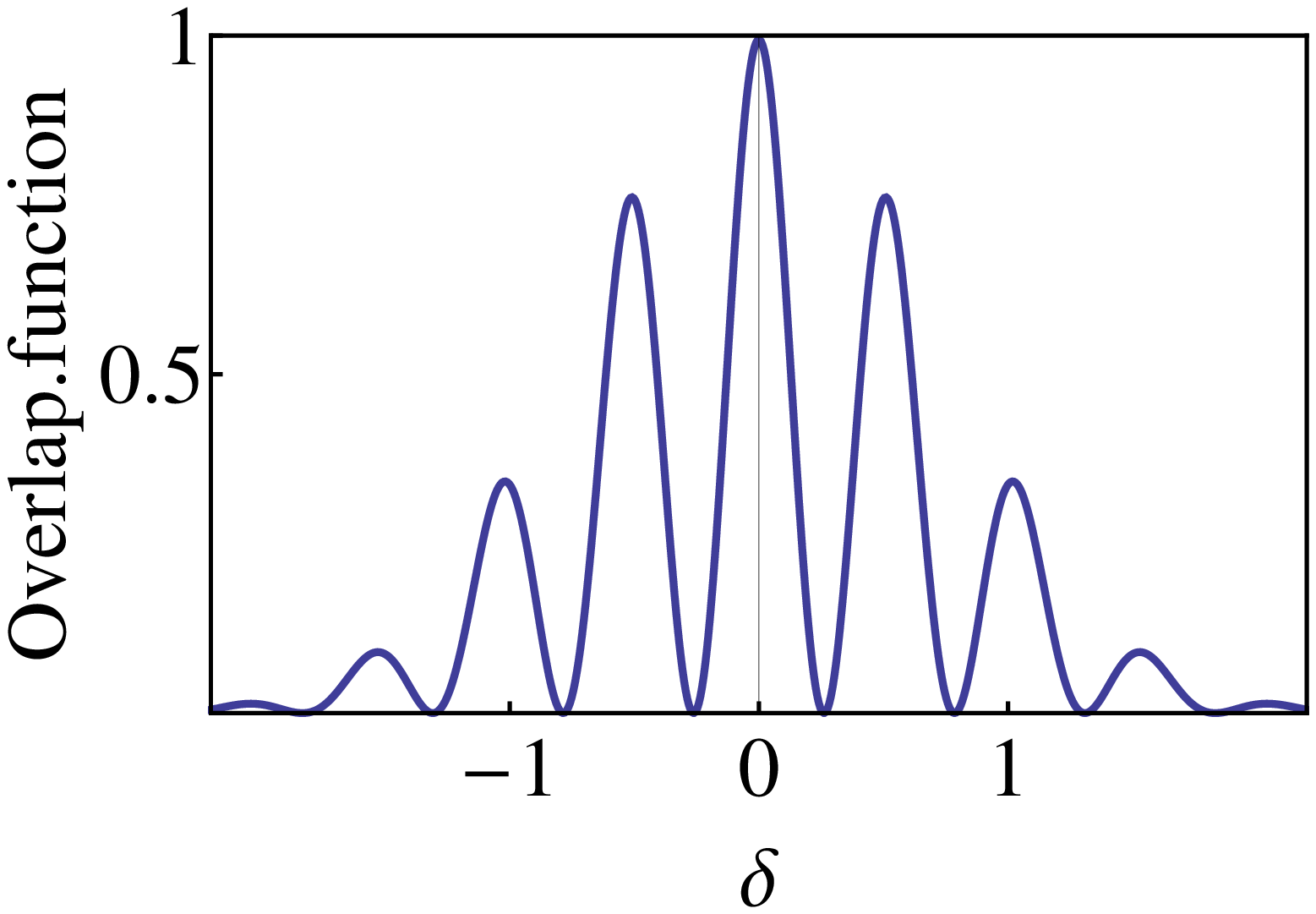}}}}%
\caption{The overlap function of Eq.(\ref{ov}) is plotted against $\delta$, when $I\approx0$, $a=(\cos{\phi}+ i \sin\phi)/\sqrt{2}$ and $b=(\cos{\phi}- i \sin\phi)/\sqrt{2}$ with $\phi=\pi/2.02$. The values of the relevant parameters are $x_{0} = 6$, $
\eta = 1$, $p_{0} = 0.01$.} 
\end{figure}
\begin{figure}[h]
{\rotatebox{0}{\resizebox{6.0cm}{5.0cm}{\includegraphics{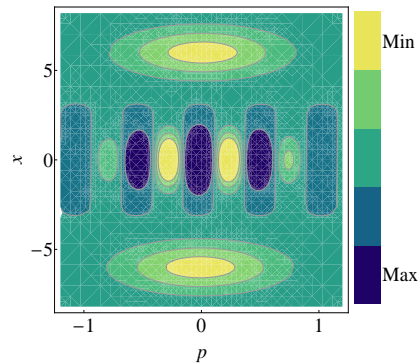}}}}%
\caption{The plot of Wigner function in phase space when $I\approx0$. The other parameters are same as in Fig. 1.  The zeros in the momentum axis exist due to the interference in phase space.}%
\end{figure}
\begin{figure}[h]
{\rotatebox{0}{\resizebox{6.0cm}{5.0cm}{\includegraphics{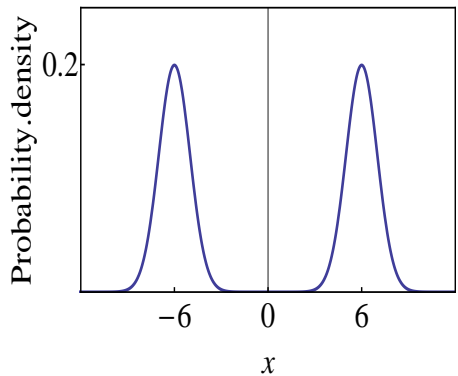}}}}%
\caption{The position distribution corresponding to Eq.(\ref{cat}) is plotted, when $I\approx0$. The values of the relevant parameters are the same as in Fig.1.}%
\end{figure}
\section{Illustrations} 
We now rigorously examine the regimes, where the sub-Planck structure and weak value emerge. For this we consider $a=(\cos{\phi}+ i \sin\phi)/\sqrt{2}$ and $b=(\cos{\phi}- i \sin\phi)/\sqrt{2}$ with $\phi=\pi/2.02$, leading to the imaginary weak value $(\sigma_{x})_{w}=-i\tan\phi$. In the context of weak measurement the imaginary weak value gives the eccentric shift in the same space the weak coupling is introduced. In our example, the shift appears in the position space. 

It is seen that, when the inner product $I$ between $	\psi_{\pm}(x)$ is zero, the oscillatory behavior of the overlap function given by Eq.(4) is explicit, as seen in Fig.1 and the corresponding presence of interference in phase space is depicted in Fig. 2, through the relevant Wigner function. Since this is the condition of strong measurement, there is no weak value in this case; the position distribution points at the eigenvalues as can be seen in Fig. 3.

Next we  consider the situation when  inner product $I$ between $	\psi_{\pm}(x)$ is one.   In this case, the overlap function is not oscillatory (see, Fig. 4) displaying no interference in phase space (Fig. 5). In this case, the position distribution $|\Psi_{c}(x)|^{2}$ depicted in Fig. 6 is peaked at the place corresponding to the imaginary part of the weak value of the observable $\hat{O}= \hat{\sigma}_{x}$.

\begin{figure}[h]
{\rotatebox{0}{\resizebox{6.0cm}{5.0cm}{\includegraphics{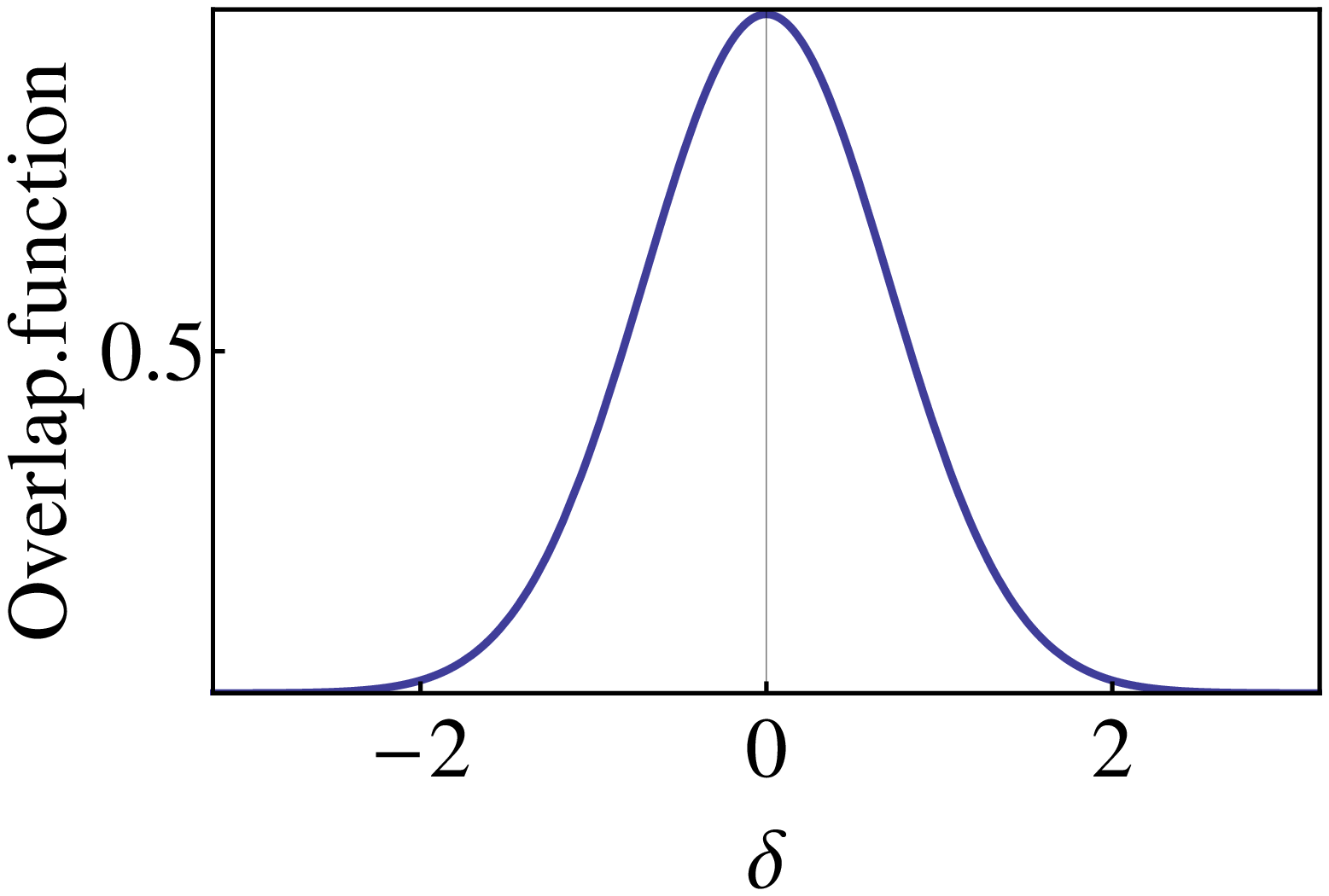}}}}%
\caption{The overlap function of Eq.(\ref{ov}) is plotted, when $I\approx1$; for $a=(\cos{\phi}+ i \sin\phi)/\sqrt{2}$ and $b=(\cos{\phi}- i \sin\phi)/\sqrt{2}$ with $\phi=\pi/2.02$. The values of the relevant parameters are  $x_{0} = 0.0001$, $\eta = 1$, $p_{0} = 0.001$.}%
\end{figure}

\begin{figure}[h]
{\rotatebox{0}{\resizebox{6.0cm}{5.0cm}{\includegraphics{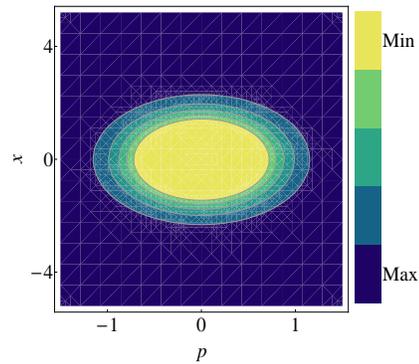}}}}%
\caption{The Wigner function corresponding to Eq.(1) shows no zeros, when $I\approx1$. The parameters are same as in Fig. 4.}%
\end{figure}

\begin{figure}[h]
{\rotatebox{0}{\resizebox{6.0cm}{5.0cm}{\includegraphics{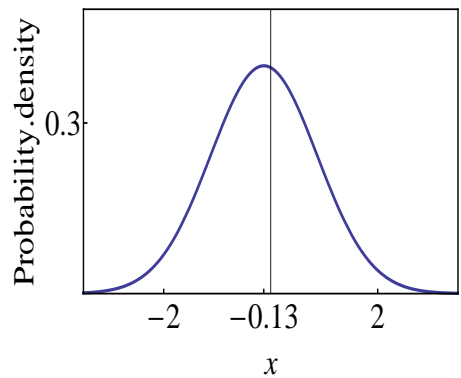}}}}%
\caption{The associated position distribution of Eq.(\ref{cat}) is plotted, when $I\approx 1$. It is peaked at $-0.13$ corresponding to the imaginary weak value $-i \tan\phi$. The values of the relevant parameters are are same as in Fig. 4.}%
\end{figure}

\section{Summary and Conclusions}
In summary, we  demonstrated that the notions of sub-Planck structure (that enables Heisenberg-limited sensitivity in quantum metrology) and weak measurement can be connected using  the cat state. Both the phenomena rely on interference effect between the two Gaussian wave functions comprising the cat state. We showed that this cat state in position space can be identified as the post-selected meter states in AAV weak measurement scenario. The inner product between the constitutent states plays the crucial role in choosing the domains, as to where the sub-Planck structure manifests and the weak value appears. We show that the sub-Planck structure and weak measurement appear in two opposite regimes, for two extreme values of the inner product, $0$ and $1$ respectively. In other words, the presence of sub-Planck structure in phase space for a given cat state obliterate the possibility of obtaining the imaginary weak value and vice versa.

Our work thus provides an interesting connection between two novel measurement scenarios, and further studies in this issue would be worthwhile by taking different types of quantum systems. It may also be interesting to study this issue for various type of constituent states, such as, non-Gaussian and orbital angular momentum states. In future, we would like to explore the usefulness of other quantum states for weak measurement, whose relevance for Heisenberg-limited measurement has been already demonstrated \cite{roy}. One may also like to understand the utility of entangled states for this purpose. A careful analysis of the experimental demonstration of weak measurement and sub-Planck structure, should be carried out to establish their relevance for both these measurement scenarios simultaneously.

\appendix*\section{}

Here, we provide the details of how the state given in Eq.(\ref{cat}) can be obtained as a post-measurement meter state in position representation, concerning the weak measurement of the observable $\hat{O}= \hat{\sigma}_{x}$ for a specific pre- and post-selection. Given that the particle is in a definite state, the AAV weak measurement procedure consists of two different SG setups, first one for introducing weak coupling (without detection) and a subsequent strong measurement setup for post-selecting a specific ensemble (see Fig.7). We consider, a beam of spin 1/2 particles, passing through the SG setups. 
The  initial total wave function is $\Psi_{in} =\psi_{0}({\bf x})|\chi_{in}\rangle$, where $\left\vert \chi_{in}\right\rangle= a |\uparrow\rangle_{x}+ b|\downarrow\rangle_{x}$ is the pre-selected state and $\psi_{0}({\bf x})$ is the spatial part, taken to
be a Gaussian wave packet peaked at the entry point (${\bf x}=0$) of the first SG
at $t=0$:
\begin{equation}
\psi_{0}\left(  \mathbf{x}\right)  =\frac{1}{{(2\pi\eta^{2})}^{{3}/{4}}}%
\exp\left(  -\frac{\mathbf{x}^{2}}{4\eta^{2}}+i\frac{p_{y}y}{\hbar}\right),
\label{initpack}%
\end{equation}
where $\eta$ is the initial width of the wave packet.

\begin{figure}[tb]
{\rotatebox{270}{\resizebox{8.0cm}{10.0cm}{\includegraphics{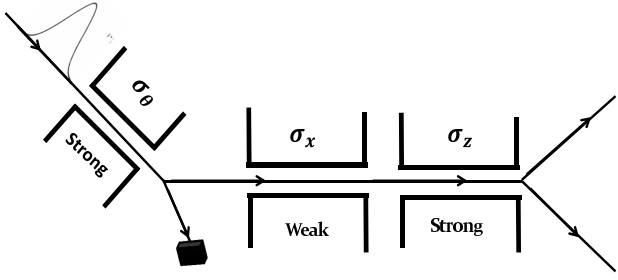}}}}%
\caption{A series of Stern-Gerlach setup for implementing weak measurement of the spin
operator $\hat{\sigma}_{x}$. The three SG setups account, respectively, for the state
preparation, the weak measurement, and the post selection .}
\end{figure}

The wave packet moves along the $+y$ axis with the initial momentum $p_{y}$ (see Fig.\ 7). The
inhomogeneous magnetic field $\mathbf{B}=(Bx,0,0)$ \footnote {This form of magnetic field is unphysical, as it does not satisfy the Maxwell
equation $\nabla.\mathbf{B}=0$. We need at least another component to make it
divergence free. If one takes, $\mathbf{B}=(B_{0}+Bx,0,-Bz)$, where $B_{0}$
is a large uniform magnetic field satisfying the condition $B_{0}>>B\delta$,
then one can  neglect the effect of $z-$component of the magnetic
field  \cite{home07}.}
is directed along the $x$-axis and is confined between
$y=0$ and $y=d$. The interaction Hamiltonian is $H_{I}=\mu\widehat{\sigma
}.\mathbf{B}$, where $\mu$ is the magnetic moment of the neutron. As
the wave packet propagates through the SG magnet, in addition to the $+y$ axis
motion, the particles gain momentum along $\pm\ x$-axis due to the interaction
of their spins with the field. The time evolved total wave function at $\tau$
(transit time of the peak of the wave packet within the SG magnetic field
region) after the interaction of spins with the SG magnetic field is given by
\begin{align}
\Psi\left(  {\bf x},\tau\right)   & =\exp\left(  {-\frac{iH_{i}\tau}{\hbar
}}\right)  \Psi({\bf x},t=0)\nonumber\\
& = a \psi_{+}(\mathbf{x},\tau)\otimes\left\vert \uparrow\right\rangle
_{x}+ b \psi_{-}(\mathbf{x},\tau)\otimes\left\vert \downarrow\right\rangle
_{x},\label{tevolved}
\end{align}
where the device states $\psi_{+}\left(  \mathbf{x},\tau\right)  $ and
$\psi_{-}\left(  \mathbf{x},\tau\right) $ are the two components of the
spinor, $\psi=\left(
\begin{array}
[c]{c}%
\begin{array}
[c]{c}%
\psi_{+}\\
\psi_{-}%
\end{array}
\end{array}
\right)  $, which satisfy the Pauli equation. Note that Eq.(\ref{tevolved}) is an entangled state between position and
spin degrees of freedom. The reduced density matrix of the system in the
$x$-basis representation can be written as
\begin{equation}
\rho_{s}=\left(
\begin{array}
[c]{cc}%
|a_{1}|^{2} & a_{1}a_{2}^{\ast} I\\
a_{1}^{\ast} a_{2} I^{\ast} & |a_{2}|^{2}
\end{array}
\right) \label{rdm}.%
\end{equation}
Here $I$ is the overlap:%
\begin{equation}
I=\int_{v}\psi_{+}^{\ast}\left(  \mathbf{x},\tau\right)  \psi_{-}\left(
\mathbf{x},\tau\right)  d^{3}x,\label{idef}%
\end{equation}
that quantifies the weakness of the measurement. The inner product $I$
is in general complex; here in our case $I$ is always real and positive.
The values of $I$ can range from $0$ to $1$, depending upon the choices of
the relevant parameters, such as, the strength of the magnetic field ($B$), the width of the
initial wave packet ($\eta$) and the transit time($\tau$) through the field region
within SG setup. We calculate the analytical expressions of
$\psi_{+}\left(  \mathbf{x},\tau\right)  $ and $\psi_{-}\left(
\mathbf{x},\tau\right)  $ by solving the relevant Schr\"{o}dinger equations.

The two-component Pauli equation for $\psi_{+}$ and $\psi_{-}$ can be
written as
\begin{align}
i\hbar\frac{\partial\psi_{+}}{\partial t} &  =-\frac{\hbar^{2}}{2m}\nabla
^{2}{\psi}_{+}+\mu bx\psi_{+},\label{decoupled}\\
i\hbar\frac{\partial\psi_{-}}{\partial t} &  =-\frac{\hbar^{2}}{2m}\nabla
^{2}{\psi}_{-}-\mu bx\psi_{-}.%
\end{align}
The solutions of the above two equations at $t=\tau$, upon exiting the SG are,
\begin{align}
\psi_{+}\left(  \mathbf{x};\tau\right)   &  =\frac{1}{\left(  2\pi\eta^{2}\right)  ^{\frac{3}{4}}}\exp\left[  -\frac{z^{2}+(y-\frac{p_{y}\tau
}{m})^{2}+(x-\frac{p_{x}^{\prime}\tau}{2m})^{2}}{4\eta^{2}}\right]
\nonumber\label{sol1}\\
&  \times\exp\left[  i\left\{  -\Delta+\left(  y-\frac{p_{y}\tau}%
{2m}\right)  \frac{p_{y}}{\hbar}+\frac{p_{x}^{\prime}x}{\hbar}\right\}
\right]
\end{align}%
and
\begin{align}
\psi_{-}\left(  \mathbf{x};\tau\right)   &  =\frac{1}{\left(  2\pi\eta
^{2}\right)  ^{\frac{3}{4}}}\exp\left[  -\frac{z^{2}+(y-\frac{p_{y}\tau
}{m})^{2}+(x+\frac{p_{x}^{\prime}\tau}{2m})^{2}}{4\eta^{2}}\right]
\nonumber\label{sol2}\\
&  \times\exp\left[  i\left\{  -\Delta+\left(  y-\frac{p_{y}\tau}%
{2m}\right)  \frac{p_{y}}{\hbar}-\frac{p_{x}^{\prime}x}{\hbar}\right\}
\right],
\end{align}
where $\Delta=\frac{{p_{x}^{\prime}}^{2}\tau}{6 m \hbar}%
$,~~$p_{x}^{\prime}=\mu B\tau$, and the spreading of the wave packet is
neglected throughout the evolution.

Here $\psi_{+}\left(  \mathbf{x},\tau\right)  $ and $\psi_{-}\left(
\mathbf{x},\tau\right)  $, representing the spatial wave functions at $\tau$,
correspond to the spin states $\left\vert \uparrow\right\rangle _{x}$ and
$\left\vert \downarrow\right\rangle _{x}$ respectively, with the average
momenta $\langle\widehat{p}\rangle_{\uparrow}$ and $\langle\widehat{p}%
\rangle_{\downarrow}$, where $\langle\widehat{p}\rangle_{\uparrow\downarrow
}=(\pm p_{x}^{\prime},p_{y},0)$. Within the magnetic field the particles gain
the same magnitude of momentum $p^{\prime}_{x}=\mu B\tau$, but the directions
are such that the particles with eigenstates $|\uparrow\rangle_{x}$ and
$|\downarrow\rangle_{x}$ get the drift along $+x$-axis and $-x$-axis
respectively, while the $y$-axis momenta remain unchanged.

From these expressions of $\psi_{+x}\left(  \mathbf{x},\tau\right)
$ and $\psi_{-x}\left(  \mathbf{x},\tau\right)  $, it is straightforward to compute the inner product $I$:
\begin{equation}
I=\exp\left(  -\frac{\mu^{2}B^{2}\tau^{4}}{8m^{2}\eta^{2}}-\frac{2\mu
^{2}B^{2}\tau^{2}\eta^{2}}{\hbar^{2}}\right), \label{inp}%
\end{equation}
which explicitly depends upon the choices of the parameters $B$, $\eta$ and
$\tau$. Eq.(\ref{inp}) is the same as Eq.(\ref{ip}), if $x_{0}=p^{\prime}_{x} \tau/2m$ and $p_{0}=p^{\prime}_{x}$ is substituted.

Now, after emerging from this non-ideal SG magnet, the particles represented by
the entangled state given by Eq. (\ref{tevolved}) enter into another SG setup,
where a strong measurement is to be performed and the particles are post-selected
in a specific spin state.

For this purpose, we consider immediately after the wave packet exits
the weak measurement SG, a subsequent strong measurement of the spin observable
$\widehat{\sigma}_{z}$ and select those particles having the state $|\uparrow\rangle_{z}$. The final normalized post-selected state can be written, by integrating out the other degrees of freedoms except $x$, as
\begin{equation}
\Psi_{Post}({x})\mathbf{=} _{z}\langle\uparrow|\Psi({x}, \tau)\rangle= \mathcal{N}\left(a\psi_{+x}({x}%
,\tau)+ b\psi_{-x}({x},\tau)\right),\label{xyzpost}%
\end{equation}
where 

\begin{equation}
\psi_{\pm x}({x},\tau)=\frac{1}{\left(  2\pi\eta^{2}\right)
^{\frac{1}{4}}}\exp\left[  -\frac{(x\mp\frac{p_{x}^{\prime}\tau}{2m})^{2}%
}{4\eta^{2}}\pm i\frac{p_{x}^{\prime}x}{\hbar}\right] \label{xpsi}%
\end{equation}
Now, if we substitute $x_{0}=p^{\prime}_{x} \tau/2m$ and $p_0=p^{\prime}_{x}$, Eq. (\ref{xpsi}) is the same as Eq. (\ref{catpm}). In such a case, $\Psi_{Post}(x)$ in Eq. (\ref{xyzpost}) is the cat state $\Psi_{c}(x)$ in Eq.(\ref{cat}). As discussed earlier $I\approx 1$ is the weak measurement limit. 
From Eq.(\ref{inp}), we can see that for a fixed $\eta$ one can
choose the other parameters $x_{0}$ and $p_{0}$, such that $I\approx1$ is obtained.\\

{\it Acknowledgements:} AKP acknowledges the support from JSPS Postdoctoral Fellowship for Foreign Researcher and Grant-in-Aid for JSPS fellows no. 24-02320.

\end{document}